# Pressure-induced superconductivity in epitaxially-stabilized $Pr_3Ni_2O_7$ films

Motoki Osada[1,2,*], Chieko Terakura[3], Hsiao-Yi Chen[4], Akiko Kikkawa[3], Masamichi Nakajima[3], Ryoma Asai[2], Jean-Baptiste Morée[3], Yusuke Nomura[4,5], Ryotaro Arita[3,6], Yoshinori Tokura[2,3,7], Atsushi Tsukazaki[1,2,4,*]

[1] *Quantum-Phase Electronics Center (QPEC), The University of Tokyo, Hongo, Tokyo, Japan*

[2] *Department of Applied Physics, The University of Tokyo, Hongo, Tokyo, Japan*

[3] *RIKEN Center for Emergent Matter Science (CEMS), Wako, Saitama, Japan*

[4] *Institute for Materials Research (IMR), Tohoku University, Sendai, Miyagi, Japan*

[5] *Advanced Institute for Materials Research (WPI-AIMR), Tohoku University, Sendai, Miyagi, Japan*

[6] *Department of Physics, The University of Tokyo, Hongo, Tokyo, Japan*

[7] *Tokyo College, The University of Tokyo, Hongo, Tokyo, Japan*

## Abstract

The discovery of high critical-temperature $T_c$ superconductivity in $La_3Ni_2O_7$ under high pressure has led to a rapid expansion of the $T_c$ range through lanthanide *Ln* substitution, and to ambient-pressure superconductivity in strained thin films, yet the exploration of new bilayer nickelates remains strongly constrained by thermodynamic stability. Beyond the difficulty of synthesis of bulk single-crystals, here we report on the pressure-induced high-$T_c$ superconductivity in the epitaxially-stabilized $Pr_3Ni_2O_7$ thin films. While the $Pr_3Ni_2O_7$ films exhibit insulating behaviour at ambient pressure regardless of ozone-annealing treatment, they show *T*-linear metallic transport and superconductivity reaching an onset $T_c$ of 66 K and zero-resistance at nearly 40 K at 22 GPa. Furthermore, $Nd_3Ni_2O_7$, with the smaller rare-earth ion Nd, can also be stabilized, however, superconductivity is not observed in the measured pressure range. Epitaxial stabilization enables us to examine the dependence of $T_c$ and critical pressure $P_c$ of superconductivity on the *Ln* ion in $Ln_3Ni_2O_7$ (*Ln* = La, Pr, Nd). These results suggest that a higher $P_c$ is required for smaller *Ln* ions, consistent with trends observed in bulk studies of *Ln* substitution. This study demonstrates that epitaxial stabilization is a powerful technique to further expand the family of superconducting bilayer nickelates.

---

osada@ap.t.u-tokyo.ac.jp, tsukazaki@ap.t.u-tokyo.ac.jp

The discovery of high critical-temperature $T_c$ superconductivity at ~78 K in bulk $La_3Ni_2O_7$ under high pressure of ~14 GPa has opened a new frontier of superconductivity[1]. Pressure-induced structural evolution, which drives a symmetry change from orthorhombic *Amam* to *Fmmm* or tetragonal *I4/mmm*, is discussed as a main trigger to induce superconductivity[2–9]. The resulting modification of the Ni $d_{3z^2-r^2}$ and Ni $d_{x^2-y^2}$ orbitals and its connection to superconductivity have been discussed theoretically[10–20]. This framework of bilayer model highlights a distinct interplay between lattice symmetry and superconductivity in the bilayer nickelates. Subsequently, ambient-pressure superconductivity was achieved in epitaxial thin films of $La_3Ni_2O_7$ and $(La,Pr)_3Ni_2O_7$ grown on $SrLaAlO_4$ substrates[21,22], where it is considered that in-plane compressive strain ($\varepsilon \sim -2\%$) straightens the Ni–apical oxygen ($O_{ap}$)–Ni bond angle toward 180° (ref.[23]). In addition, another $LaAlO_3$ substrates ($\varepsilon \sim -1.2\%$) is applicable to host superconductivity both at ambient pressure with onset $T_c$ ($T_c^{onset}$) ~ 10 K after ozone-annealing treatments[24] and under high pressure with $T_c^{onset}$ ~ 60 K at 20 GPa (ref.[25]). These films facilitate experimental approaches to achieving a comprehensive understanding of the superconductivity in bilayer nickelates by spectroscopic investigations of the electronic states and by detail physical properties measurements[26–28]. Alongside theoretical considerations, experimental progress in material discovery and property evaluation are advancing rapidly in both bulk and thin-film systems.

Regarding recent materials advances, superconductivity under high-pressure has been reported in *A*-site *Ln*-substituted $La_{3-x}Ln_xNi_2O_7$ (*Ln* = Pr, Nd, Sm): the maximum isovalent substitution levels reach $x = 2.67$, 2.13, and 1.43 for Pr, Nd, and Sm, respectively[1–3]. In these systems, the maximum $T_c^{onset}$ increases from ~78 K for $La_3Ni_2O_7$ to nearly 96 K for $La_{1.57}Sm_{1.43}Ni_2O_{7-\delta}$ as the effective lattice constants decrease by substituting *Ln* with small ionic radius[1–3]. This trend suggests that chemical pressure provides a promising strategy for enhancing $T_c$. However, there is upper limit of *Ln* substitution in $La_{3-x}Ln_xNi_2O_7$ due to thermodynamic stability and structural symmetry breaking. Following this perspective, fully *A*-site-replaced $Ln_3Ni_2O_7$ is an alternative candidate for hosting superconductivity. Indeed, theoretical studies have extensively investigated structural symmetry[29] and whether $T_c$ increases[30] or decreases[31] with small ionic radii. Despite these theoretical interests, $Ln_3Ni_2O_7$ compounds remain largely unexplored in both bulk and thin-film forms due to thermodynamic instability of the bilayer stacking. In the Ruddlesden-Popper compounds, the window of stable synthesis conditions for the bilayer stacking is extremely narrow, and the existence of other Ruddlesden-Popper phases easily leads to phase separation[32]. In this context, epitaxial stabilization technique plays a crucial role to achieve $Ln_3Ni_2O_7$ compounds in thin-film form[33,34]. In this study, we tried to fabricate $Pr_3Ni_2O_7$ thin films, overcoming the difficulty of synthesizing this phase as a bulk crystal. Upon applying hydrostatic pressure to the film, we observed the emergence of superconductivity with an $T_c^{onset}$ of 66 K and zero resistance near 40 K at 22 GPa.

## Instability of $Pr_3Ni_2O_7$ compared to $La_3Ni_2O_7$

We considered the instability of $Pr_3Ni_2O_7$ in terms of symmetry and formation energies of Ruddlesden-Popper phases. Following the theoretical examination of the lattice symmetry with the Goldschmidt tolerance factor, $t\ (A) = (r_A + r_O)/[\sqrt{2}(r_{Ni} + r_O)]$ (ref.[29]), $t = 1$ corresponds to an ideally packed perovskite structure with Ni-O-Ni bond angle of 180°, whereas deviations from unity indicate increasing symmetry breaking. In $Ln_3Ni_2O_7$, two crystallographically distinct *A*-site positions exist as depicted in Fig. 1a: *Ln* (1) within the perovskite blocks and *Ln* (2) located between the $NiO_6$ bilayers, corresponding to coordination numbers of 12 (XII) and 9 (IX), respectively. Although the perovskite blocks contain *Ln* (1) sites, we use the ionic radius $r_{A,IX}$ (ref.[35]) for the modified tolerance-factor of the bilayer nickelates $t'\ (A)$ and subsequent discussion. Figure 1b shows the modified tolerance factor $t'\ (A)$ as a function of $r_A$, which systematically decreases from La to Sm. A systematic theoretical investigation of structural properties of bilayer nickelates proposed a structural transition from *Cmcm* symmetry in $La_3Ni_2O_7$ and $Pr_3Ni_2O_7$ to $Cmc2_1$ symmetry in $Nd_3Ni_2O_7$ and $Sm_3Ni_2O_7$ at ambient pressure[29]. Notably, $Pr_3Ni_2O_7$ lies at the boundary of this predicted structural transition, as indicated by the shaded region in Fig. 1b. This border suggests an increased level of geometric instability, which may explain why bulk $Pr_3Ni_2O_7$ has remained largely unexplored compared with $La_3Ni_2O_7$.

We examine whether first-principles calculations capture the experimentally known difficulty of synthesizing bulk $Ln_3Ni_2O_7$ (*Ln*327) with *Ln* = Pr compared with *Ln* = La. As a simplified energetic reference, we here compare the density functional theory (DFT) total energy of *Ln*327 with those of $Ln_2NiO_4$ (*Ln*214) and $LnNiO_3$ (*Ln*113), both of which are well-established nickelate compounds[36–38] (see Methods for details). According to the relative energetic positions for *Ln* = La and Pr shown in Fig. 1c, we find that Pr327 is higher in energy than La327, indicating less stability of Pr327. Although this comparison does not establish thermodynamic stability in the strict convex-hull sense, the result provides insight into the experimental difficulty of synthesizing bulk $Pr_3Ni_2O_7$. Along with the further understanding of the thermodynamic stability by construction of the convex-hull system, it is worth pursuing the synthesis of $Pr_3Ni_2O_7$ in thin films via epitaxial strain beyond the difficulty of bulk synthesis.

## Epitaxial Stabilization of $Pr_3Ni_2O_7$ in thin-film form

To employ an epitaxial strain, we prepared $Ln_3Ni_2O_7$ thin films ($Ln$ = La, Pr, Nd) on $LaAlO_3$ (001) substrates by pulsed-laser deposition (Fig. 1a), followed by ex-situ ozone-annealing treatment (see Methods for details). The X-ray 2theta-omega scan revealed diffraction peaks corresponding to the $Pr_3Ni_2O_7$ 00*l* diffractions on $LaAlO_3$ (001) substrate (Fig.1d). To probe the in-plane symmetry, we performed in-plane phi-scans of the $Pr_3Ni_2O_7$ 206 and $LaAlO_3$ 101 diffractions (Fig. 1e) showing four-fold symmetry. In Fig. S1a shows a reciprocal space map (RSM) around $LaAlO_3$ 103 and $Pr_3Ni_2O_7$ 11$\underline{17}$ peaks, revealing that the $Pr_3Ni_2O_7$ film is fully locked to the substrate in-plane lattice constant of 3.79 Å and resultantly an out-of-plane (*c*-axis) lattice constant is 20.51 Å. These structural characterization with X-ray diffraction characterizes the epitaxial stabilization of the $Pr_3Ni_2O_7$ phase with four-fold symmetry, which is distinct from the two-fold symmetry of bulk $Ln_3Ni_2O_7$ at ambient pressure.

Local cross-sectional imaging by scanning transmission electron microscopy (STEM) enables direct visualization of the atomic structure in thin films. Figures 1f and g show a high-angle annular dark-field (HAADF) STEM images of $Pr_3Ni_2O_7/LaAlO_3$ films, projected along the [110] direction for $Pr_3Ni_2O_7$ and the [100] direction for $LaAlO_3$. In the $Pr_3Ni_2O_7$ film, Pr columns appear brightest, while Ni columns are relatively less intense. The observed atomic arrangement indicates that $NiO_6$ bilayer (BL) stacking is dominant, characterized by an in-plane shift of 0.5 unit-cell (u.c.) every two $NiO_6$ octahedra. Upon closer inspection in Fig. 1f, monolayer (ML) $NiO_6$ stacking is locally observed. This feature is more clearly identified as dark-blue region with orange arrows in the strain-mapping image shown in Fig. S1b and S1c, which maps the lattice strain along the out-of-plane (*c*-axis) direction $\varepsilon_{yy}$ of the $PrNiO_3$ perovskite blocks and overlays it with the HAADF-STEM image (see Methods for details)[39]. These microscopic characterizations, along with X-ray characterizations, consistently indicate that the $Pr_3Ni_2O_7$ film is stabilized by a predominantly bilayer-stacked structure.

## Pressure-induced Superconductivity in $Pr_3Ni_2O_7$

Figure 2a shows the temperature-dependent resistivity of as-grown and ozone-annealed $Pr_3Ni_2O_7$ films at ambient pressure. Ozone-annealing reduces the resistivity by approximately a factor of four compared with the as-grown film; however, the film remains insulating. A clear resistive kink is observed around 140 K, below which the insulating behavior is more pronounced. This feature looks similar in a comparable temperature region to that reported for $La_3Ni_2O_{7-\delta}$ (refs.[40,41]) and $La_3Ni_2O_7$ (ref.[42]) and in $La_3Ni_2O_7$ thin-film without ozone-annealing[25], which may be associated with the formation of a density-wave state.

Notably, at hydrostatic pressure of 22 GPa, a clear superconducting transition emerges in ozone-annealed $Pr_3Ni_2O_7$ film. As shown in the left inset of Fig. 2b, the resistivity decreases monotonically upon cooling, followed by a pronounced resistive downturn at low temperatures. In particular, $T$-linear metallic transport is observed below ~140 K (main panel). We define the onset superconducting transition temperature, $T_c^{onset}$, as the intersection point of linear extrapolations of the normal-state resistivity and the transition region, yielding $T_c^{onset}$ = 66 K. This value is higher than the $T_c^{onset}$ values reported in the nickelate thin films at ambient pressure at roughly 40 K (refs.[21,22]) and the $T_c^{onset}$ of 60 K observed for $La_3Ni_2O_7$ film measured at 20 GPa (ref.[25]) and 22 GPa by using the identical high-pressure experimental setup (Fig. S2). Furthermore, the observation of zero resistance below 37 K (right inset) provides compelling evidence for superconductivity in $Pr_3Ni_2O_7$.

## Resistivity of $Ln_3Ni_2O_7$ films under Pressure

Figure 3 presents the pressure-dependent resistivity of ozone-annealed $Pr_3Ni_2O_7$, $La_3Ni_2O_7$, and $Nd_3Ni_2O_7$ films on $LaAlO_3$ substrates (structural characterization by X-ray diffraction in Fig. S3 for $La_3Ni_2O_7$ and $Nd_3Ni_2O_7$ films, and HAADF-STEM images in Fig. S4 for $Nd_3Ni_2O_7$ films). To ensure high hydrostaticity, we employed a cubic-anvil cell with a liquid pressure-transmitting medium, a technique well verified in our previous work[25] (see Methods for details). Firstly, at low pressures, $Pr_3Ni_2O_7$ remains insulating behavior up to 16 GPa, whereas the low-temperature insulating state is progressively suppressed with increasing pressure, suggesting the enhancement of in-plane hopping via lattice squeezing. Upon further compression, $Pr_3Ni_2O_7$ undergoes an insulator-to-metal crossover and displays metallic temperature dependence at 20 GPa, followed by the emergence of superconductivity with $T_c^{onset}$ of ~60 K. At 22 GPa, the metallic behavior is further enhanced and the $T_c^{onset}$ increases to 66 K. These results demonstrate a pressure-driven insulator-to-metal crossover in $Pr_3Ni_2O_7$ with emergence of superconductivity. Secondly, ozone-annealed $La_3Ni_2O_7$ film exhibits metallic temperature dependence, and a resistive downturn emerges around 10 K at 4 GPa in Fig. 3b. This critical pressure $P_c$ is substantially lower than that of as-grown $La_3Ni_2O_7$ films without ozone-annealing, for which superconductivity has been reported to appear above ~16 GPa with an onset temperature of ~50 K (ref.[25]). Although the effect of ozone-annealing treatment is still under debate[43,44], the oxygen filling in the lattice strongly correlates to the in-plane electrical conduction via both Ni $d_{3z^2-r^2}$ and Ni $d_{x^2-y^2}$ bands. In fact, the ozone-annealed $La_3Ni_2O_7$ film shows metallic transport at ambient pressure in contrast to insulator in the as-grown film. Based on this enhancement of metallic conduction, the observed small $P_c$ may be consistent with recent observation of ambient-pressure superconductivity (~10 K) in ozone-annealed $La_2PrNi_2O_7/LaAlO_3$ films[24]. Thirdly, ozone-annealed $Nd_3Ni_2O_7$ film also exhibits enhanced metallic transport with increasing pressure in Fig. 3c. Despite this trend, no superconducting transition was observed up to 20 GPa. Given the increased $P_c$ in recent studies on *Ln*-substituted $(La,Ln)_3Ni_2O_7$ bulks[2,3], $Nd_3Ni_2O_7$ films may require pressures above 20 GPa. A theoretical study considering lattice symmetry in bulk predicts that a structural transition from *Cmcm* symmetry to *I4/mmm* occurs at 10–20 GPa for $La_3Ni_2O_7$, 20–30 GPa for $Pr_3Ni_2O_7$, and 30–40 GPa for $Nd_3Ni_2O_7$ (ref.[29]). For reasons possibly related to the smaller ionic radius and/or magnetic nature of Nd, no superconducting transition is observed, and the system instead exhibits metallic transport. Finally, epitaxial stabilization enables us to examine the dependence of resistivity $\rho_{xx}$ and $T_c$ on the *Ln* ion in $Ln_3Ni_2O_7$ (*Ln* = La, Pr, Nd). As shown in Fig. 3d, the smaller *A*-site ions, Nd and Pr, exhibit comparable behaviour at $P$ = 20 GPa, despite their higher resistivity at ambient pressure. In addition, the $T_c$ at $P$ = 20 GPa is summarized in Fig. 3e, showing comparable values of La and Pr. Further studies on applying high pressure via diamond anvil cell is effective to examine whether high-$T_c$ superconductivity appears in $Nd_3Ni_2O_7$ film or not.

## Phase Diagram: Thin films vs. Bulk crystals

Figure 4 summarizes the link between the $T_c$ and $P_c$ in temperature–pressure (*T*–*P*) phase diagrams of $Ln_3Ni_2O_7$ thin films on $LaAlO_3$ substrates (left) and their bulk-crystal counterparts (right). Viewed together, the thin-film and bulk phase diagrams reveal overall consistency on the appearance of high $T_c$ at high pressure region. Based on this consistent trend, clear distinctions emerge; (1) superconductivity in thin films appears at 4 GPa, which is lower than approximately 8 GPa in bulk crystals. (2) Highest $T_c$ value in thin films ~66 K is lower than that in bulk crystals of *Ln*-substituted $(La,Ln)_3Ni_2O_7$ ~ 96 K. Specifically in $La_3Ni_2O_7$, it is plausible that the epitaxial strain contributes complementary to reduce in $P_c$ as role of chemical pressure. The regulation of the optimal lattice by both chemical pressure, achieved through composition tuning, and hydrostatic pressure, plays a crucial role for the appearance of high-$T_c$. The attainable $T_c$ in thin films under high pressure will be improved by reduction of structural inhomogeneity relating to Ruddlesden-Popper stacking and/or oxygen filling. This combined perspective provides a unified framework for tuning superconductivity in bilayer nickelates.

## Conclusion

The realization of the $Pr_3Ni_2O_7$ phase by epitaxial stabilization, together with the observation of superconductivity at 66 K under hydrostatic pressure, establishes a viable route to finding new bilayer nickelates that are otherwise constrained by thermodynamic instability. The higher $T_c$ compared with the La-based counterpart highlights the role of chemical pressure introduced by reducing the *A*-site ionic radius. Consistent with this trend, theoretical and experimental studies have shown that smaller rare-earth ions can enhance magnetic interactions, suggesting that chemical pressure broadly influences superconductivity in layered nickelates. Notably, superconductivity in bilayer $Pr_3Ni_2O_7$ emerges at comparatively lower pressures (~20 GPa) highlighting bilayer $Pr_3Ni_2O_7$ as a particularly accessible platform. Our work provides an effective route to expanding experimental access to other *Ln*-based bilayer nickelates, which remain of significant interest for further investigation.

## Methods

**Sample preparation:** $Ln_3Ni_2O_7$ films ($Ln$ = La, Pr, and Nd) were fabricated on single-crystalline $LaAlO_3$ (001) substrates using pulsed-laser deposition. $Ln_3Ni_2O_7$ targets were prepared by mixing precursor powders ($Ln_xO_y$ and NiO) and sintering them at 1200°C for 12 hours, repeated twice. The polycrystalline targets were ablated by a KrF excimer laser (wavelength 248 nm). Substrates were pre-annealed at 650°C in an oxygen partial pressure of $1 \times 10^{-6}$ Torr to obtain an atomically flat surface. During growth, the substrate temperature was fixed at 650°C and the oxygen partial pressure was 200–300 mTorr. The laser fluence was 0.7–1.0 J/cm$^2$ and a repetition rate was 4 Hz. The thickness of $La_3Ni_2O_7$, $Pr_3Ni_2O_7$, and $Nd_3Ni_2O_7$ were approximately 11 nm, 10 nm, and 7 nm, respectively. $SrTiO_3$ capping layer (~1 u.c.) was deposited on the top of films. After the deposition, the sample is transferred ex-situ to ozone-annealing treatment apparatus (Samco UV-1). Ozone-annealing was carried out at 240–300 °C for a total of approximately 100 minutes under a 500 sccm ozone flow. The films were characterized using x-ray diffraction (XRD) techniques with Cu K$\alpha$ source ($\lambda$ = 1.5406 Å). Crystal structure depicted in Fig. 1a were visualized using the VESTA software[45].

**Scanning transmission electron microscopy:** To prepare cross-sectional lamellas, we employed a focused ion beam (FIB) lift-out procedure using Thermo Fisher Scientific Versa 3D and Quanta 3D systems. Scanning transmission electron microscopy (STEM) was conducted using JEM-ARM200F STEMCorr. STEM imaging was performed at an accelerating voltage of 200 kV. Strain mapping images were analyzed using the Strain++ software based on the geometric phase analysis (GPA) method[39].

**Electrical transport measurement:** The measurements of the temperature dependent resistivity $\rho_{xx}(T)$ were measured using a six-point geometry with Au wire bonded on Au electrodes (30 nm-thick) in physical properties measurement system (PPMS, Quantum Design, Inc.). Au electrode was deposited at room temperature by electron-beam evaporation.

**High-pressure electric resistivity measurement:** The thin-film samples were cut into dimensions of 500–1000 μm in lateral size, and the substrates were mechanically polished to reduce their thickness to 0.3 mm (see Fig. S5 for details). Gold wires were bonded to the samples in a van der Pauw geometry using silver paste to ensure reliable electrical contacts. Thin-film samples were mounted in a Teflon capsule and MgO gasket. Pressure was induced using a cubic-anvil cell with a liquid pressure-transmitting medium (Daphne oil 7575, Idemitsu Kosan Co., Ltd.) to apply hydrostatically and to carefully suppress the occurrence of inhomogeneity in the film. Press-loading speeds are: 1.6 ml/min in 0–1 GPa, 1.0 ml/min in 1–4 GPa, 1.0 ml/min in 4–8 GPa, 0.8–1.0 ml/min in 0.8–12 GPa, 0.6–0.8 ml/min in 12–16 GPa, and 0.4 ml/min in 16–22 GPa. The $\rho_{xx}(T)$ curves at $P = 0$ GPa were observed in physical properties measurement system (PPMS) (Quantum Design, Inc.). The measurements of the temperature dependent resistivity $\rho_{xx}(T)$ under high-pressure were performed under various hydrostatic pressure of 1, 4, 8, 12, 16, 20, and 22 GPa. The temperature dependence of resistance was measured at 1 μA using Keithley 2182A nanovoltmeters and Keithley 6221 source meters in delta mode. Measurements were conducted over a temperature range from 292 K to 4.2 K.

**Theoretical calculations based on density functional theory:** Total energy calculations of fully optimized atomic structures for bulk *Ln*327, *Ln*113 and *Ln*214 (*Ln* = La, Pr) were performed using the projector augmented-wave method, as implemented in the Vienna Ab initio Simulation Package (VASP)[46]. The exchange-correlation potential is treated within Perdew-Burke-Ernzerhof (PBE) generalized gradient approximation[47] while the Pr 4*f*-orbitals are treated as frozen core. A plane-wave cutoff energy of 850 eV is used to converge the variation in total energy down to 10 meV for all compounds. The Brillouin zone was sampled using Monkhorst–Pack ***k***-point meshes of 10×10×12 for *Ln*327, 10×5×8 for *Ln*113, and 10×10×6 for *Ln*214. Since the synthesis temperature is above the magnetic transition temperature, we focus on the nonmagnetic states for all compositions and on the high-symmetry *I*4/*mmm* structure for *Ln*327.

## Data availability

The data that support the findings of this study are available from the corresponding author upon request.

## Acknowledgements

The authors thank M. Kawasaki and M. Nakamura for technical supports and fruitful discussion. STEM observations were made with the cooperation of Y. Kodama and K. Hayasaka of Analytical Research Core for Advanced Materials, Institute for Materials Research, Tohoku University. A part of this work was supported by Tohoku University in MEXT Advanced Research Infrastructure for Materials and Nanotechnology in Japan (Grant No. JPMXP1225TU1066), the GIMRT

Program of the Institute for Materials Research, Tohoku University (Grant No. 202505-CRKEQ-0052), Toyota Riken Scholar Program by Toyota Physical and Chemical Research Institute, the RIKEN TRIP initiative (RIKEN Quantum, Advanced General Intelligence for Science Program, Many-body Electron Systems), and JSPS KAKENHI (Grant Nos. JP23K13663, JP24H00190, JP25K00015, JP25H01246, JP25H01250, and JP25H01252). Y.N. acknowledges support from JSPS KAKENHI (Grant Nos. JP23H04869, JP23K03307, and JP25H01506) and MEXT as "Program for Promoting Researches on the Supercomputer Fugaku" (Project ID: JPMXP1020230411).

## Author contributions

M.O. and A.T. conceived the project. M.O. fabricated and characterized nickelate films. M.O., C.T., A.K., M.N., Y.T., and A.T. performed the high-pressure measurement and analysis. H.-Y.C., R.A., J.-B.M., Y.N., and R.A. conducted theoretical calculation. M.O. and A.T. wrote the manuscript with input from all the authors.

## Competing interests

The authors declare no competing interests.

## References

1. Sun, H. *et al.* Signatures of superconductivity near 80K in a nickelate under high pressure. *Nature* **621**, 493–498 (2023).
2. Wang, N. *et al.* Bulk high-temperature superconductivity in pressurized tetragonal $La_2PrNi_2O_7$. *Nature* **634**, 579–584 (2024).
3. Li, F. *et al.* Bulk superconductivity up to 96 K in pressurized nickelate single crystals. *Nature* **649**, 871–878 (2026).
4. Hou, J. *et al.* Emergence of high-temperature superconducting phase in pressurized $La_3Ni_2O_7$ crystals. *Chin. Phys. Lett.* **40**, 117302 (2023).
5. Zhang, Y. *et al*. High-temperature superconductivity with zero resistance and strange-metal behaviour in $La_3Ni_2O_{7-\delta}$. *Nat. Phys.* **20**, 1269–1273 (2024).
6. Wang, G. *et al.* Pressure-induced superconductivity in polycrystalline $La_3Ni_2O_{7-\delta}$. *Phys. Rev. X.* **14**, 011040 (2024).
7. Dong, Z. *et al*. Visualization of oxygen vacancies and self-doped ligand holes in $La_3Ni_2O_{7-\delta}$. *Nature* **634**, 579–584 (2024).
8. Yang, J. *et al*. Orbital-dependent electron correlation in double-layer $La_3Ni_2O_{7-\delta}$. *Nat. Commun.* **15**, 4373 (2024).

9. Chen, X. *et al.* Electronic and magnetic excitations in $La_3Ni_2O_7$. *Nat. Commun.* **15**, 9597 (2024).
10. Nakata, M., Ogura, D., Usui, H. & Kuroki, K. Finite-energy spin fluctuations as a pairing glue in systems with coexisting electron and hole bands. *Phys. Rev. B* **95**, 214509 (2017).
11. Maier, T. A., Mishra, V., Balduzzi, G. & Scalapino, D. J. Effective pairing interaction in a system with an incipient band. *Phys. Rev. B* **99**, 140504(R) (2019).
12. Luo, Z., Hu, X., Wang, M., W´u, W., and Yao, D.-X. Bilayer two-orbital model of $La_3Ni_2O_7$ under pressure, *Phys. Rev. Lett.* **131**, 126001 (2023).
13. Zhang, Y., Lin, L.-F., Moreo, A., and Dagotto, E. Electronic structure, dimer physics, orbitalselective behavior, and magnetic tendencies in the bilayer nickelate superconductor $La_3Ni_2O_7$ under pressure, *Phys. Rev. B* **108**, L180510 (2023).
14. Christiansson, V., Petocchi, F., and Werner, P. Correlated electronic structure of $La_3Ni_2O_7$ under pressure, *Phys. Rev. Lett.* **131**, 206501 (2023).
15. Lechermann, F., Gondolf, J., Bötzel, S., and Eremin, I. M. Electronic correlations and superconducting instability in $La_3Ni_2O_7$ under high pressure, *Phys. Rev. B* **108**, L201121 (2023).
16. Sakakibara, H., Kitamine, N. Ochi, M. & Kuroki, K. Possible High $T_c$ Superconductivity in $La_3Ni_2O_7$ under High Pressure through Manifestation of a Nearly Half-Filled Bilayer Hubbard Model. *Phys. Rev. Lett.* **132**, 106002 (2024).
17. Lu, C., Pan, Z., Yang, F., and Wu, C. Interlayer-coupling-driven high-temperature superconductivity in $La_3Ni_2O_7$ under pressure, *Phys. Rev. Lett.* **132**, 146002 (2024).
18. Yang, Q.-G., Wang, D., and Wang, Q.-H. Possible $s^{\pm}$-wave superconductivity in $La_3Ni_2O_7$, *Phys. Rev. B* **108**, L140505 (2023).
19. Zhang, Y. Lin, L.-F. Moreo, A., Maier, T. A., and Dagotto, E. Structural phase transition, $s^{\pm}$wave pairing, and magnetic stripe order in bilayered superconductor $La_3Ni_2O_7$ under pressure, *Nat. Commun.* **15**, 2470 (2024).
20. Nomura, Y., Kitatani, M., Sakai, S., and Arita, R. Strong-coupling high-$T_c$ superconductivity in doped correlated band insulators. *Phys. Rev. Lett. B* **112**, L020504 (2025).
21. Ko, E. K. *et al.* Signatures of ambient pressure superconductivity in thin film $La_3Ni_2O_7$. *Nature* **638**, 935–940 (2025).
22. Zhou, G. *et al*. Ambient-pressure superconductivity onset above 40 K in $(La,Pr)_3Ni_2O_7$ films. *Nature*. **640**, 641–646 (2025).
23. Bhatt, L. *et al.* Resolving Structural Origins for Superconductivity in Strain-Engineered $La_3Ni_2O_7$ Thin Films. Preprint at arxiv.org/abs/2501.08204 (2025).
24. Tarn, Y. *et al.* Reducing the strain required for ambient-pressure superconductivity in Ruddlesden-Popper bilayer nickelates. Adv. Mater. **38**, 220724 (2026).
25. Osada, M. *et al.* Strain-tuning for superconductivity in $La_3Ni_2O_7$ thin films. *Commun. Phys.* **8**, 251 (2025).
26. Li, P. *et al.* Angle-resolved photoemission spectroscopy of superconducting $(La,Pr)_3Ni_2O_7/SrLaAlO_4$ heterostructures. *National Science Review* **12**, nwaf205 (2025).

27. Wang, B. Y. *et al.* Electronic structure of compressively strained thin film $La_2PrNi_2O_7$. Preprint at arxiv.org/abs/2504.16372 (2025).
28. Nie, Z. *et al.* Ambient-pressure superconductivity and electronic structures of engineered hybrid nickelate films. Preprint at arxiv.org/abs/2509.03502 (2025).
29. Geisler, B., Hamlin, J. J., Stewart, G. R. Hennig, R. G., and Hirschfeld, P. J. Structural transitions, octahedral rotations, and electronic properties of $A_3Ni_2O_7$ rare-earth nickelates under high pressure. *npj Quantum Materials* **9**, 38 (2024).
30. Pan, Z., Lu, C., Yang, F., and Wu, C. Effect of Rare-Earth Element Substitution in Superconducting $R_3Ni_2O_7$ under Pressure *Chin. Phys. Lett.* **41**, 087401 (2024).
31. Zhang, Y., Lin, L.-F., Moreo, A., Maier, T. A., and Dagotto, E. Trends in electronic structures and $s\pm$-wave pairing for the rare-earth series in bilayer nickelate superconductor $R_3Ni_2O_7$. *Phys. Rev. B* **108**, 165141 (2023).
32. Zhang, Z., Greenblatt, and J. B., Goodenough. Synthesis, Structure, and Properties of the Layered Perovskite $La_3Ni_2O_{7-\delta}$. *J. Solid. State Chem.* **108**, 402–409 (1994).
33. Dawley, N. M.*et al*. Thermal cconductivity of the $n$ = 1–5 and 10 members of the $(SrTiO_3)_nSrO$ RuddlesdenPopper superlatices. *Appl. Phys. Lett.* **118**, 091904 (2021).
34. Pan, G. A. *et al.* Synthesis and electronic properties of $Nd_{n+1}Ni_nO_{3n+1}$ Ruddlesden-Popper nickelate thin films. *Phys. Rev. Mater.* **6**, 055003 (2022).
35. Shannon, R. D. Revised Effective Ionic Radii and Systematic Studies of Interatomic Distances in Halides and Chalcogenides. *Acta Cryst.* A **32**, 751–767 (1976).
36. Guo, H. *et al.* Antiferromagnetic correlations in the metallic strongly correlated transition metal oxide $LaNiO_3$. *Nat. Commun.* **9**, 43 (2018).
37. Cui, Y., Luo, Z., Zhang C., Ren, Y., and Gao, Y. Electrical and optical properties of $NdNiO_3/PrNiO_3$ heterojunctions upon oxygen vacancy. *Mater. Sci. Eng.B* **298**, 116866 (2023).
38. Fernández-Díaz, M. T. *et al.* Structural and magnetic phase transitions in $Pr_2NiO_4$. *Z. Phys. B – Condensed Matter* **82**, 275–282 (1991).
39. Hÿtch, M. J., Snoeck, E. and Kilaas, R. Quantitative measurement of displacement and strain fields from HREM micrographs. *Ultramicroscopy* **74**, 131–146 (1998).
40. Shi, M. *et al.* Spin density wave rather than tetragonal structure is prerequisite for superconductivity in $La_3Ni_2O_{7-\delta}$. *Nat. Commun.* **16**, 9141 (2025).
41. Luo, J. *et al.* Microscopic Evidence of Charge- and Spin-Density Waves in $La_3Ni_2O_{7-\delta}$ Revealed by $^{139}$La-NQR. *Chi. Phys. Lett.* **42**, 067402 (2025).
42. Khasanov, R. *et al.* Pressure-enhanced splitting of density wave transitions in $La_3Ni_2O_{7-\delta}$. *Nat. Phys.* **21**, 430–436 (2025).
43. Liu, Y. *et al.* Superconductivity and normal-state transport in compressively strained $La_2PrNi_2O_7$ thin films. *Nat. Mater.* **24**, 1221–1227 (2025).
44. Dong, Z. *et al.* Interstitial oxygen order and its competition with superconductivity in $La_2PrNi_2O_{7+\delta}$. *Nat. Mater.* **24**, 1927–1934 (2025).
45. Momma K. & Izumi, F. Vesta 3 for three-dimensional visualization of crystal, volumetric and morphology data. *J. Appl. Crystallogr.* **44**, 1272–1276, (2011).

46. Kresse, G. and Furthmüller, J. Efficient iterative schemes for ab initio total-energy calculations using a plane-wave basis set. *Phys. Rev. B* **54**, 11169 (1996).
47. Perdew, J. P., Burke, K., and Ernzerhof, M. Generalized Gradient Approximation Made Simple. *Phys. Rev. Lett.* **77**, 3865 (1996).

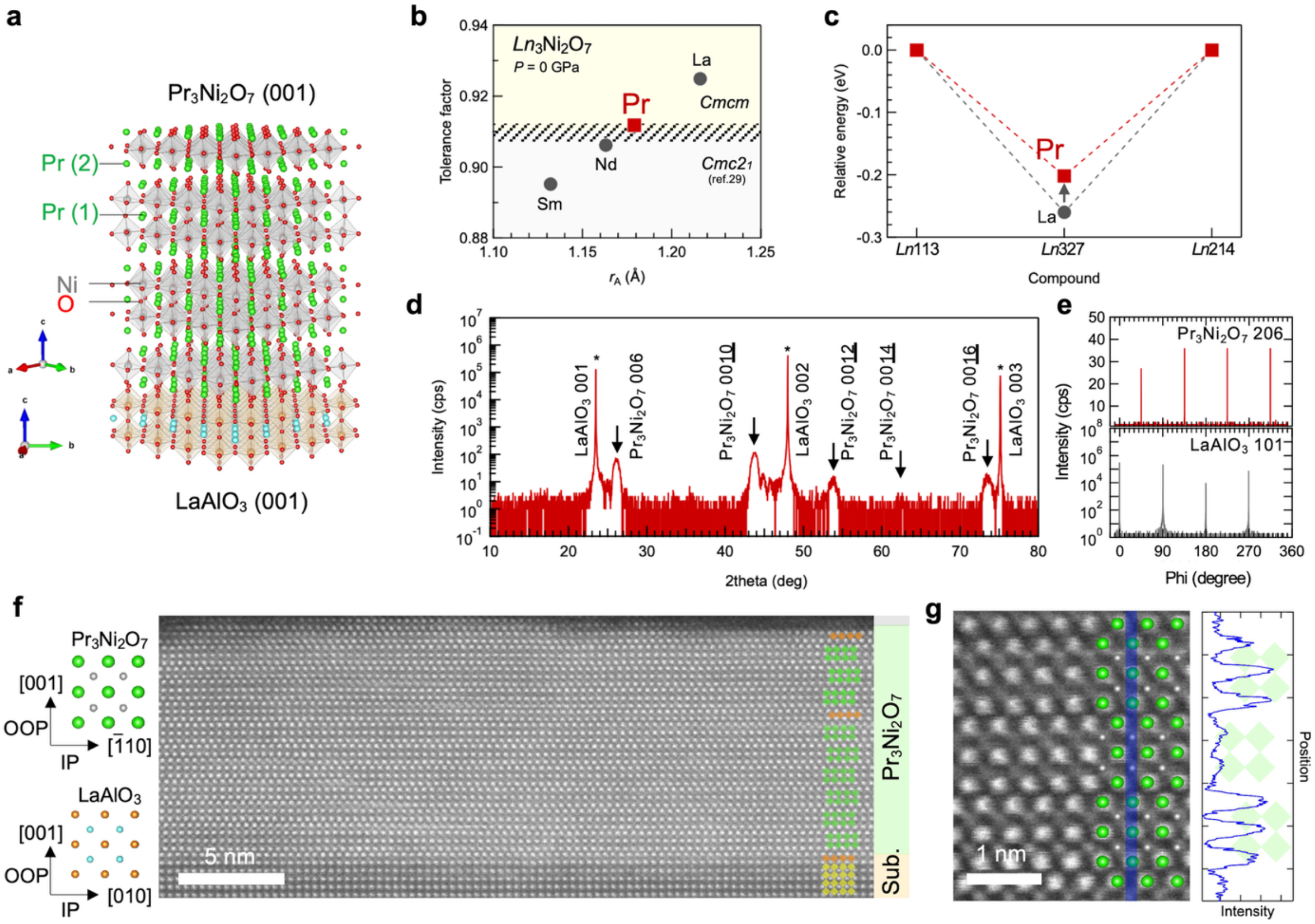

**Fig. 1| Epitaxial stabilization of $Pr_3Ni_2O_7$ thin films.** **a**, Crystal structure of $Pr_3Ni_2O_7$ on $LaAlO_3$. **b**, The modified tolerance factor for bilayer nickelates $Ln_3Ni_2O_7$ plotted as a function of the *A*-site ionic radius $r_{A,\mathrm{IX}}$. Structural transition adapted from ref.[29]. **c,** Analysis of the relative stability of $Ln_3Ni_2O_7$ (*Ln* = La, Pr) using $Ln_2NiO_4$ and $LnNiO_3$ as reference systems. The DFT energy of $Ln_3Ni_2O_7$ is compared to that of $(1\text{-}x)Ln_2NiO_4 + xLnNiO_3$ at $x = 0.5$ (see Methods for details). **d**, X-ray 2theta-omega scan of ozone-annealed $Pr_3Ni_2O_7$ thin film on $LaAlO_3$ substrate. **e**, Phi scans of $Pr_3Ni_2O_7$ 206 and $LaAlO_3$ 101. **f**, High-angle annular dark-field (HAADF)-STEM image of $Pr_3Ni_2O_7$ thin film on $LaAlO_3$ substrate taken along $Pr_3Ni_2O_7$ [110] and $LaAlO_3$ [100] axis. OOP and IP represent out-of-plane and in-plane, respectively. The schematic atomic arrangement is overlayed with yellow circles for La in $LaAlO_3$ and green and orange circles for Pr in $Pr_3Ni_2O_7$. **g**, Enlarged STEM image of bilayer structure and intensity profile along the blue line. The schematic atomic arrangement is overlayed with green circles for Pr in $Pr_3Ni_2O_7$.

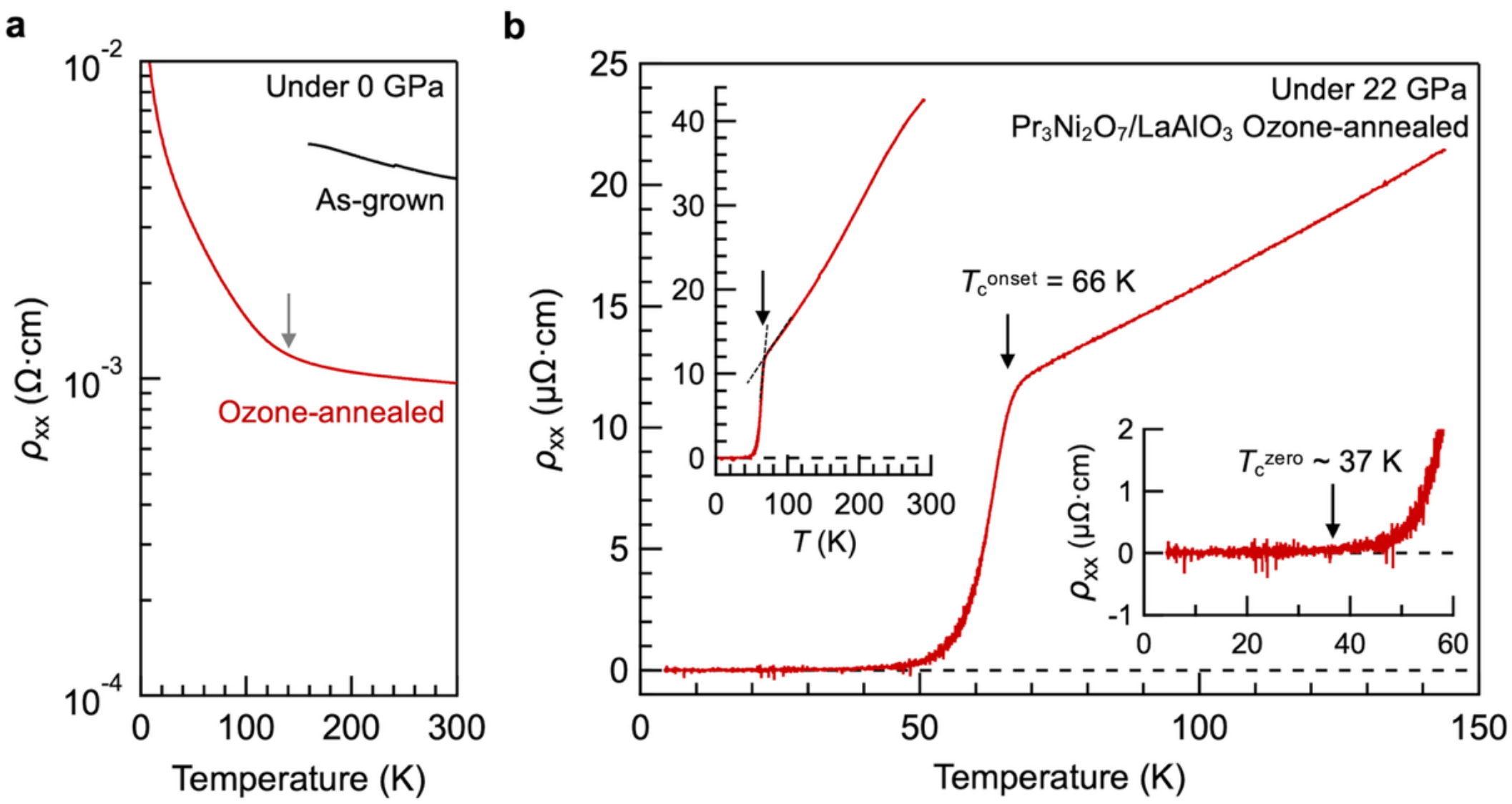

**Fig. 2 | Superconductivity of $Pr_3Ni_2O_7$ films under high pressure and insulating behaviour at ambient pressure. a**, Temperature dependence of resistivity for as-grown and ozone-annealed $Pr_3Ni_2O_7$ films at ambient pressure. Arrow indicates a resistive kink. **b**, Temperature dependence of resistivity of ozone-annealed $Pr_3Ni_2O_7$ film under 22 GPa. The left inset shows resistivity ranges up to 300 K. The two broken lines for estimation of the onset $T_c$. The right inset shows zero resistance at 37 K. Arrows indicate $T_c^{onset}$ and $T_c^{zero}$.

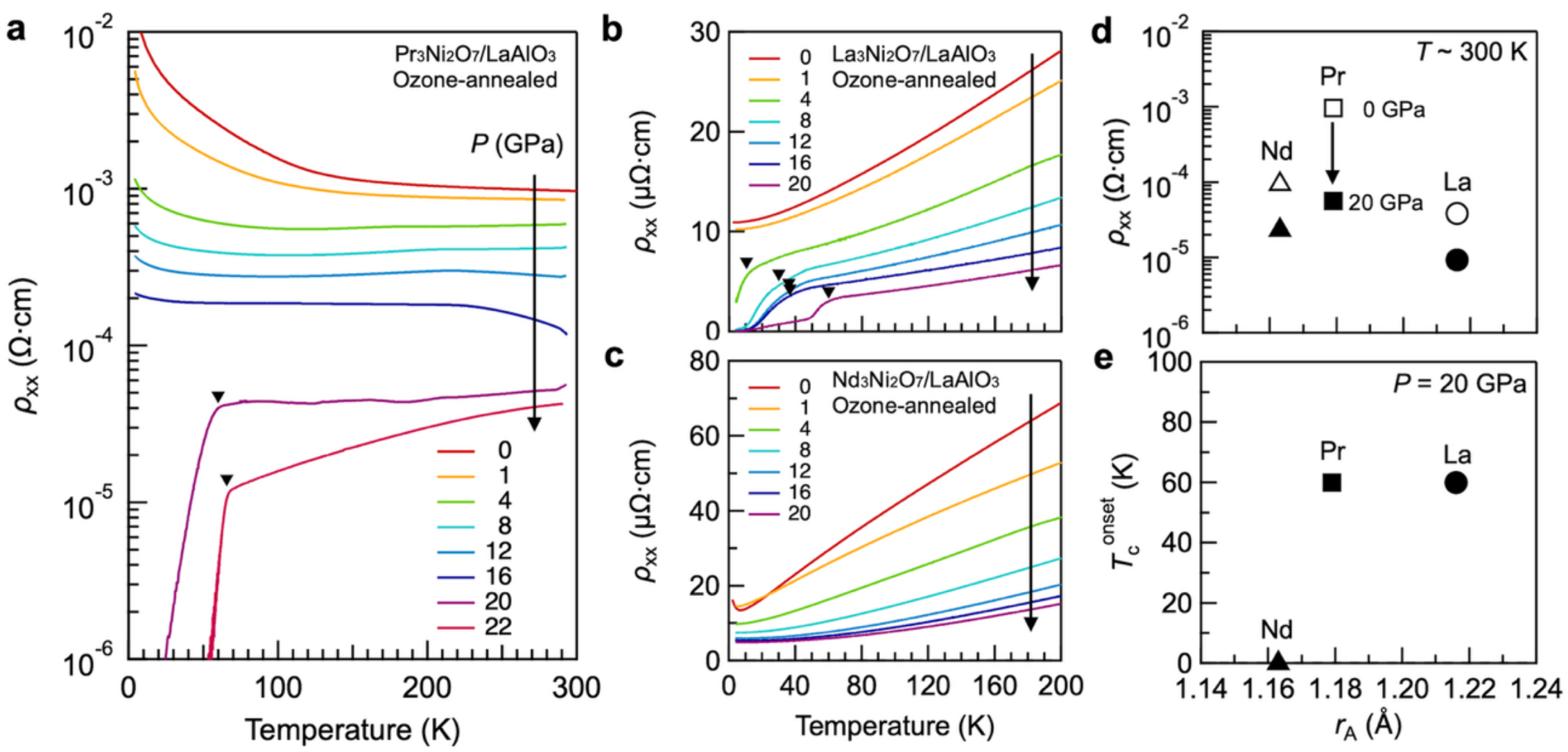

**Fig. 3 | Pressure-induced variations in temperature dependence of resistivity of ozone-annealed *Ln*$_3$Ni$_2$O$_7$ films.** Temperature dependent resistivity under various pressures for **a**, $Pr_3Ni_2O_7$, **b**, $La_3Ni_2O_7$, and **c**, $Nd_3Ni_2O_7$ films on $LaAlO_3$ substrate. **d**, resistivity changes of the $Ln_3Ni_2O_7$ films for *Ln* = La, Pr, and Nd by applying 20 GPa at $T = 300$ K. **e**, $T_c$ for $Ln_3Ni_2O_7$ films under 20 GPa.

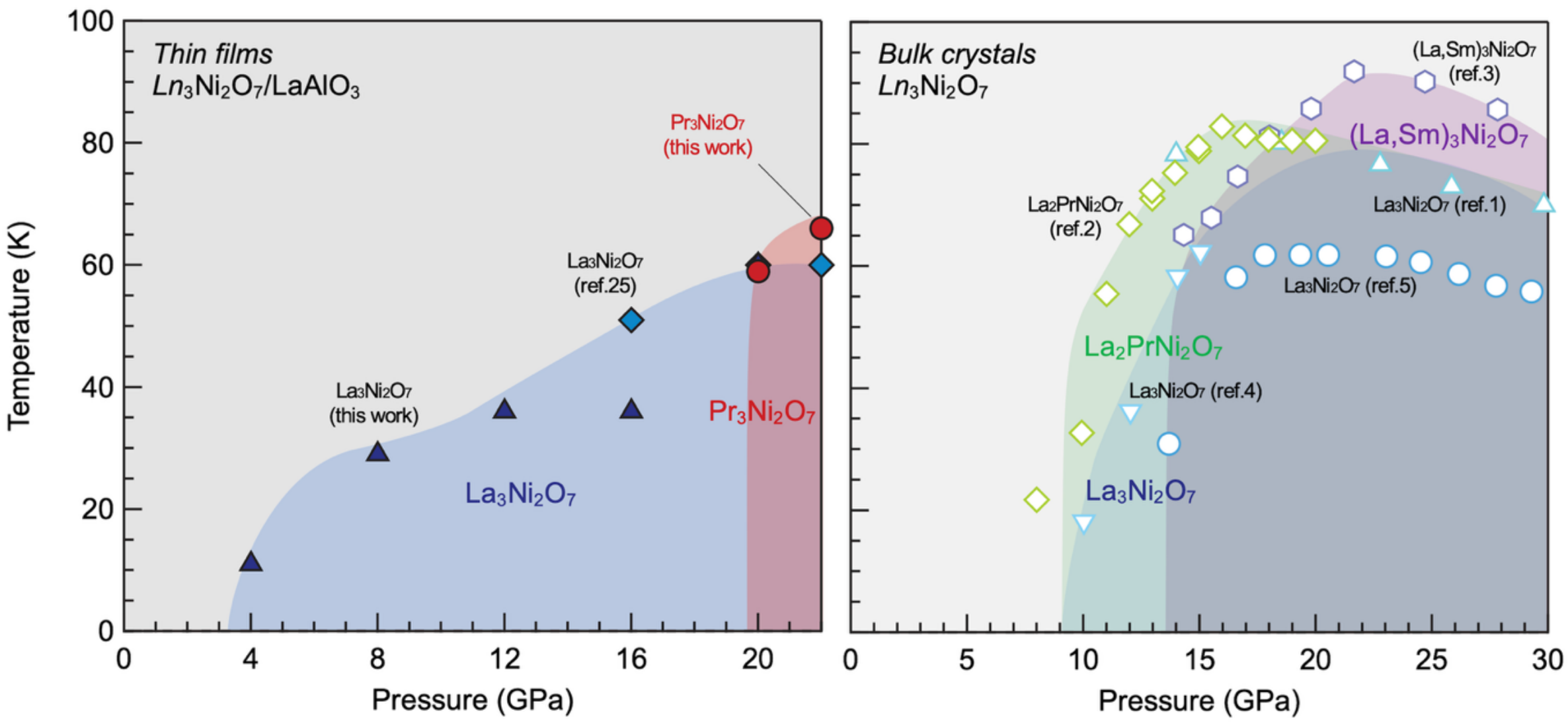

**Fig. 4 | *T–P* phase diagram of *Ln*$_3$Ni$_2$O$_7$.** *T–P* phase diagram for *Ln*$_3$Ni$_2$O$_7$/LaAlO$_3$ thin films (left) and *Ln*$_3$Ni$_2$O$_7$ bulk crystals (right). (Left) Closed circles, triangles, and diamonds represent ozone-annealed Pr$_3$Ni$_2$O$_7$/LaAlO$_3$, La$_3$Ni$_2$O$_7$/LaAlO$_3$ thin films and as-grown La$_3$Ni$_2$O$_7$/LaAlO$_3$ thin films (ref.[25]), respectively. (Right) Open circles, upward and downward triangles, diamonds, hexagons indicate La$_3$Ni$_2$O$_7$, La$_2$PrNi$_2$O$_7$, and (La,Sm)$_3$Ni$_2$O$_7$ bulk cystals (refs.[1–5]).

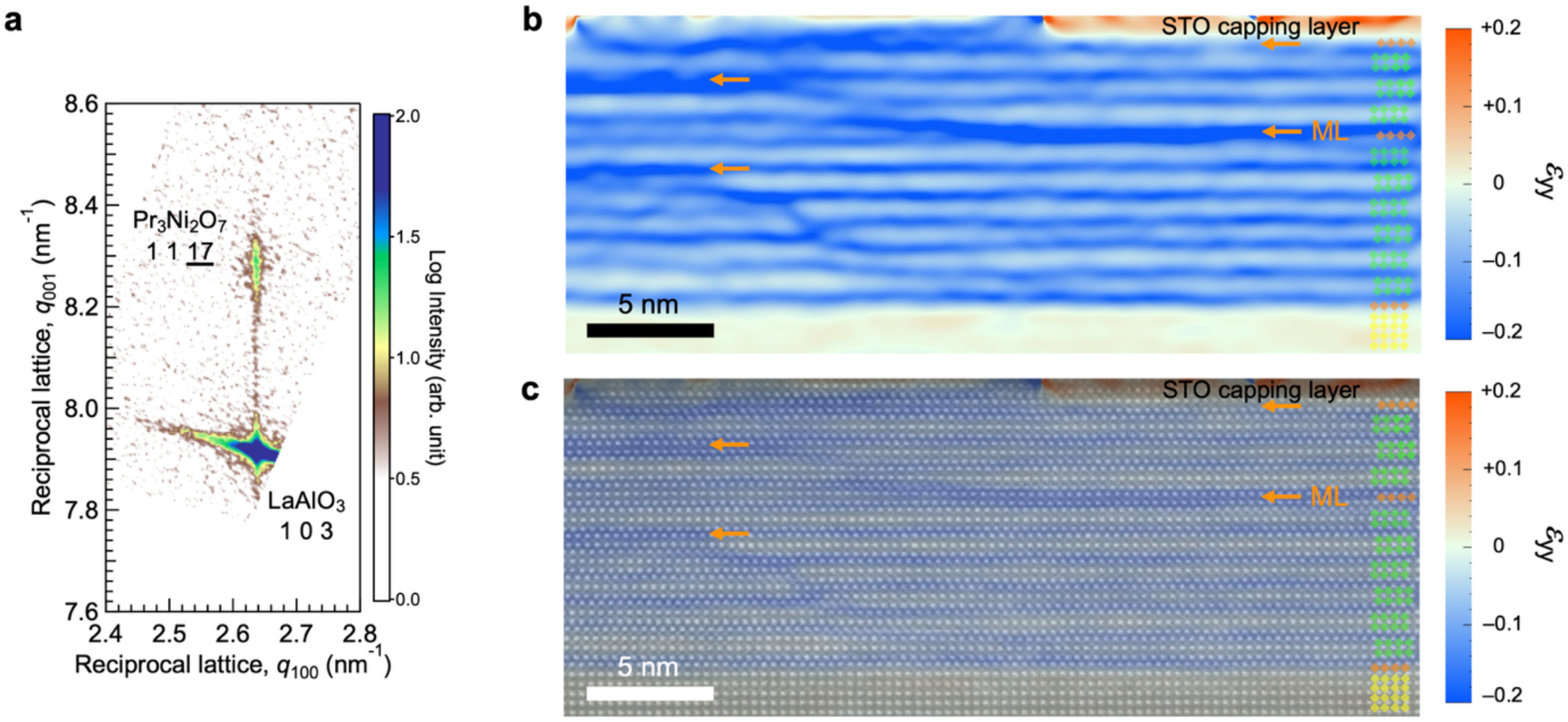

**Fig. S1 | Reciprocal space mapping (RSM) and strain matting of $Pr_3Ni_2O_7$ film. a**, RSM image of $Pr_3Ni_2O_7$ film on $LaAlO_3$ substrate. **b**, Strain map taken along $Pr_3Ni_2O_7$ [110] axis and $LaAlO_3$ [100] axis. Strain along out-of-plane direction $\varepsilon_{yy}$ is analyzed. The schematic atomic arrangement is overlayed with green circles for Pr in $Pr_3Ni_2O_7$. **c**, Strain map is overlayed on high-angle annular dark field (HAADF) image taken along $Pr_3Ni_2O_7$ [110] axis and $LaAlO_3$ [100] axis. Wide dark blue region corresponds to monolayer (ML) stacking. $SrTiO_3$ (STO) capping layer is deposited on top of the film (red region). The schematic atomic arrangement is overlayed with green circles for Pr in $Pr_3Ni_2O_7$.

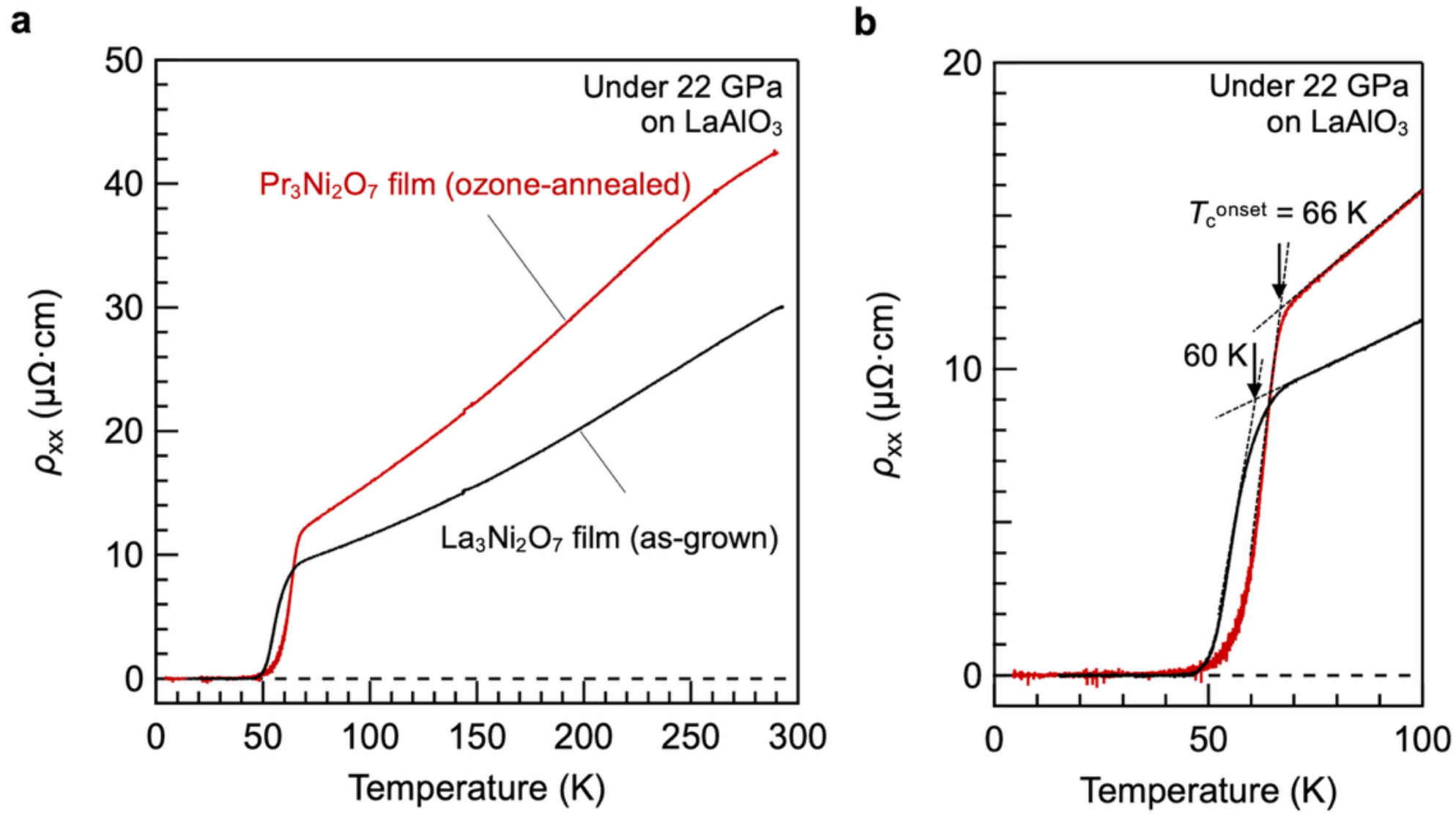

**Fig. S2 | Pressure-induced superconductivity of $La_3Ni_2O_7$ and $Pr_3Ni_2O_7$ films.** **a**, Temperature dependent resistivity of $La_3Ni_2O_7$ (as-grown) and $Pr_3Ni_2O_7$ films (ozone-annealed) on $LaAlO_3$ under 22 GPa. **b**, Resistivity below 100 K, showing onset $T_c$ of 60 K and 66 K for $La_3Ni_2O_7$ and $Pr_3Ni_2O_7$ films, respectively.

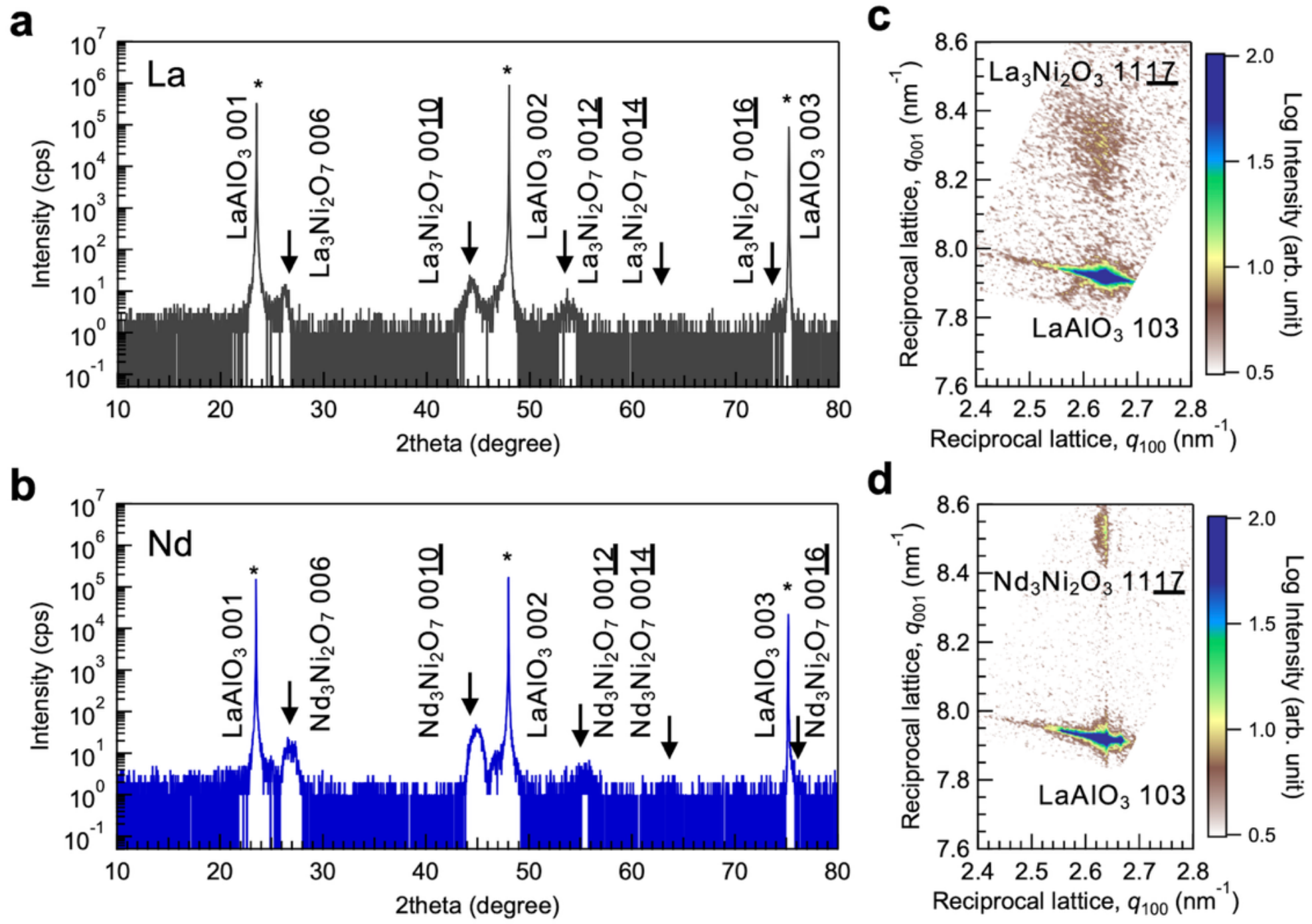

**Fig. S3 | X-ray diffraction of *Ln*$_3$Ni$_2$O$_7$ thin films. a**, 2theta-omega scan of ozone-annealed $La_3Ni_2O_7$ film on $LaAlO_3$ substrate. **b**, 2theta-omega scan of ozone-annealed $Nd_3Ni_2O_7$ film on $LaAlO_3$ substrate. **c**, Reciprocal space mapping of ozone-annealed $La_3Ni_2O_7$ film on $LaAlO_3$ substrate. **d**, Reciprocal space mapping of ozone-annealed $Nd_3Ni_2O_7$ film on $LaAlO_3$ substrate.

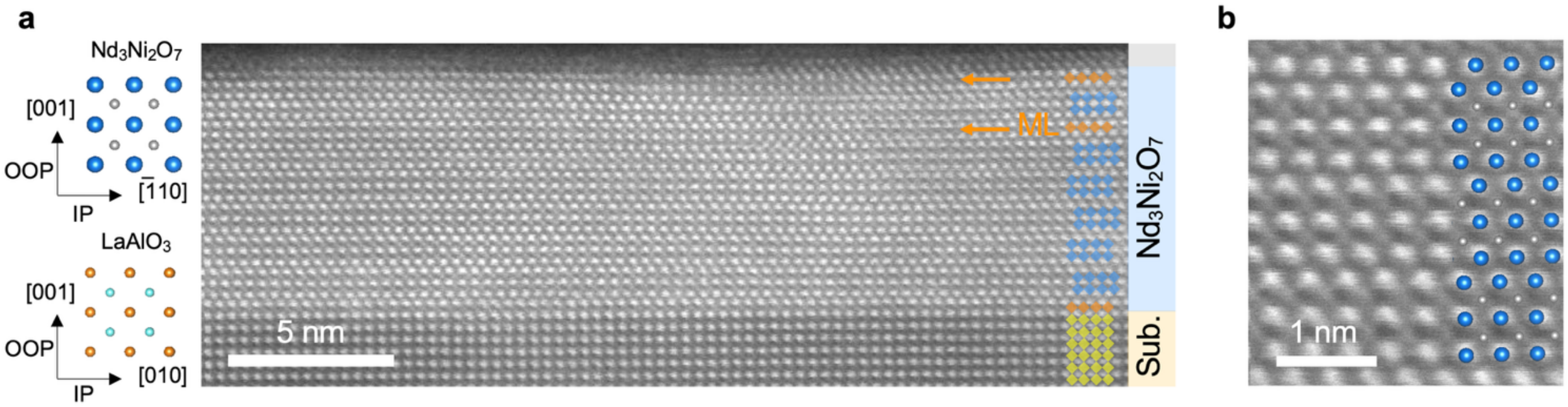

**Fig. S4 | STEM image of Nd-based bilayer nickelate $Nd_3Ni_2O_7$ films. a**, High-angle annular dark-field (HAADF)-STEM image of $Nd_3Ni_2O_7$ thin film on $LaAlO_3$ substrate taken along $Nd_3Ni_2O_7$ [110] and $LaAlO_3$ [100] axis. OOP and IP represent out-of-plane and in-plane, respectively. The schematic atomic arrangement is overlayed with blue circles for Nd in $Nd_3Ni_2O_7$. ML denotes a monolayer stacking. **b**, Enlarged STEM image of bilayer structure. The schematic atomic arrangement is overlayed with blue circles for Nd in $Nd_3Ni_2O_7$.

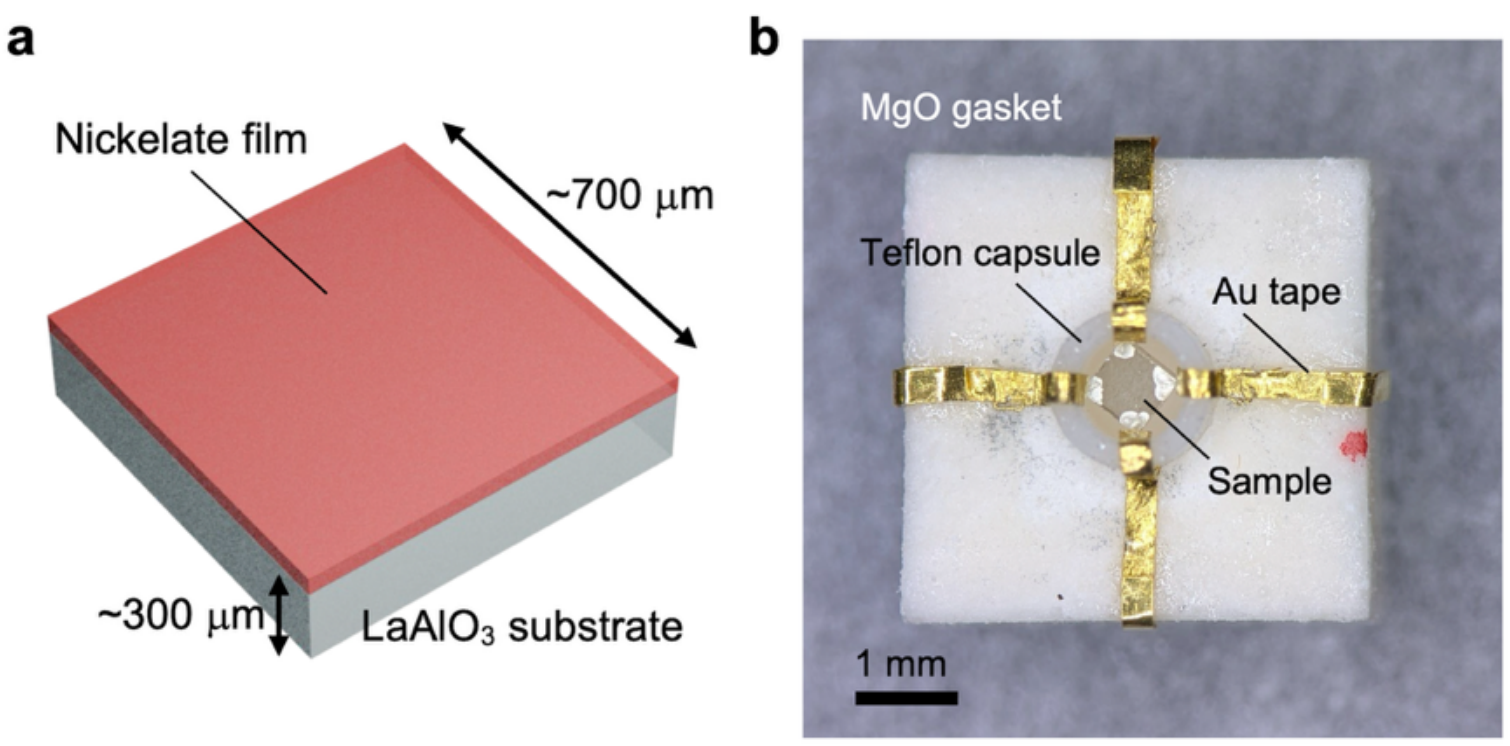

**Fig. S5 | Sample geometry for high-pressure electric resistivity measurement. a**, Schematic illustration of the thin-film sample geometry used for resistivity measurements under high pressure. **b**, Photograph of the cubic gasket employed in cubic-anvil-cell-type high-pressure experiments.